\documentclass[a4paper,fleqn,usenatbib]{mnras}

\usepackage{newtxtext,newtxmath}

\usepackage[T1]{fontenc}
\usepackage{ae,aecompl}

\usepackage{graphicx}	
\usepackage{amsmath}	
\usepackage{amssymb}	

\usepackage{tabu}
\usepackage{longtable}

\title[AGN merger enhancement at z$\sim$2]{The redshift evolution of major merger triggering of luminous AGN: a slight enhancement at z$\sim$2}

\author[T. Hewlett et al.]{
Timothy Hewlett,$^{1}$\thanks{E-mail: th51@st-andrews.ac.uk (TH)}
Carolin Villforth,$^{1,2}$
Vivienne Wild,$^{1}$
Jairo Mendez-Abreu,$^{1,3,4}$
\newauthor Milena Pawlik$^{1}$
and Kate Rowlands$^{1,5}$
\\
$^{1}$Department of Physics and Astronomy, University of St Andrews, North Haugh, St Andrews, KY16 9SS, Scotland\\
$^{2}$University of Bath, Department of Physics, Claverton Down, Bath BA2 7AY, United Kingdom\\
$^{3}$Instituto de Astrofísica de Canarias, Calle Vía Láctea s/n, E-38205 La Laguna, Tenerife, Spain\\
$^{4}$Departamento de Astrofísica, Universidad de La Laguna, E-38200 La Laguna, Tenerife, Spain\\
$^{5}$Department of Physics \& Astronomy, Johns Hopkins University, Bloomberg Center, 3400 N. Charles St., Baltimore, MD 21218, USA
}

\date{Accepted 2017 April 24. Received 2017 April 21; in original form 2016 October 06}

\pubyear{2017}

\begin{document}
\label{firstpage}
\pagerange{\pageref{firstpage}--\pageref{lastpage}}
\maketitle

\begin{abstract}

Active galactic nuclei (AGN), particularly the most luminous AGN, are commonly assumed to be triggered through major mergers, however observational evidence for this scenario is mixed. To investigate any influence of galaxy mergers on AGN triggering and luminosities through cosmic time, we present a sample of 106 luminous X-ray selected type 1 AGN from the COSMOS survey. These AGN occupy a large redshift range (0.5 < z < 2.2) and two orders of magnitude in X-ray luminosity ($\sim$10\textsuperscript{43} - 10\textsuperscript{45} erg s\textsuperscript{-1}). AGN hosts are carefully mass and redshift matched to 486 control galaxies. A novel technique for identifying and quantifying merger features in galaxies is developed, subtracting \textsc{galfit} galaxy models and quantifying the residuals. Comparison to visual classification confirms this measure reliably picks out disturbance features in galaxies. No enhancement of merger features with increasing AGN luminosity is found with this metric, or by visual inspection. We analyse the redshift evolution of AGN associated with galaxy mergers and find no merger enhancement in lower redshift bins. Contrarily, in the highest redshift bin (z$\sim$2) AGN are $\sim$4 times more likely to be in galaxies exhibiting evidence of morphological disturbance compared to control galaxies, at 99\% confidence level ($\sim$2.4$\sigma$) from visual inspection. Since only $\sim$15\% of these AGN are found to be in morphologically disturbed galaxies, it is implied that major mergers at high redshift make a noticeable but subdominant contribution to AGN fuelling. At low redshifts other processes dominate and mergers become a less significant triggering mechanism.

\end{abstract}




\section{Introduction}

Supermassive black holes (SMBHs) are believed to exist in the centres of all massive galaxies \citep{Kormendy1995}. A small proportion of these are growing, with gas accretion rates ranging from $\sim$10\textsuperscript{-4} - 10 M$_{\text{$\bigodot$}}$ year\textsuperscript{-1} and a proportionately wide range of bolometric luminosities ($\sim$10\textsuperscript{42} - 10\textsuperscript{47} erg s\textsuperscript{-1}). These are Active Galactic Nuclei (AGN) and may accrete large fractions of their mass in bursts of rapid accretion \citep{Croton2006}, requiring rapid inflow of gas from galaxy length-scales. Stripping the gas of enough angular momentum to allow for such rapid accretion, thereby powering the most luminous AGN, proves extremely challenging. Theoretical work suggests major mergers can provide the torque to displace such an overwhelming fraction of the angular momentum of the gas, allowing for the highest accretion rates onto the central black hole whilst transforming the galaxy morphology \citep{Toomre1972, Barnes1988, Barnes1991, DiMatteo2005, Cox2008}. Gas rich mergers may trigger nuclear and global starbursts \citep{Mihos1994, Mihos1996, Hopkins2006a} and major mergers disrupt the morphologies of the colliding galaxies, often exhibiting long tidal tails or shells of expelled gas and stars soon after the merger has begun. Detecting this can be challenging however, since the single new galaxy has a relaxation timescale after which morphological features of mergers fade \citep{Larson1978, Kennicutt1987, Ellison2013}. Observational evidence suggesting a link between major mergers and SMBH accretion has been mixed \citep[e.g.][]{Gabor2009, Cisternas2011, Schawinski2011, Kocevski2012, Treister2012, Villforth2013, Ellison2013, Villforth2014, Kocevski2015, Villforth2017}. Alternatively, AGN may be triggered secularly through, for example, disk instabilities \citep{Bournaud2011}, bars \citep{Knapen2000, Oh2012}, or otherwise by minor mergers \citep{Kaviraj2013}. It remains unclear whether alternatives to major merger triggering can drive several M$_{\text{$\bigodot$}}$ year\textsuperscript{-1} of gas to the central SMBHs, as is necessary to power the most luminous AGN.

The tightness of correlations between the masses of SMBHs and host galaxy masses, velocity dispersions and luminosities may imply coevolution between them \citep{Ferrarese2000, Marconi2003, Haring2004}. Major mergers are a commonly proposed mechanism for growing black holes and galaxies synchronously. Detection of connections between AGN activity and major mergers may therefore shed some light on the origins of correlations between the SMBH mass and galaxy properties, such as the M-$\sigma$ relation (the correlation between the mass of the SMBH and the velocity dispersion of the spheroid). Alternatively, these relations could be caused by stochastic processes, since black holes and star formation are both fed by a supply of cold gas.

The possibility of a connection between nuclear activity and galaxy mergers can be investigated observationally by comparing galaxy morphologies and AGN properties. \cite{Canalizo2001}, \cite{Treister2012}, \cite{Ellison2013} and \cite{Kocevski2015} find evidence of enhancement of AGN associated with mergers, or else that mergers are associated in particular with obscured AGN \citep{Kocevski2015}. However, \cite{Allevato2011} and \cite{Karouzos2014} suggest that biases can account for apparent correlations between AGN and galaxy interactions, while \cite{Grogin2005}, \cite{Gabor2009}, \cite{Cisternas2011}, \cite{Schawinski2011}, \cite{Kocevski2012}, \cite{Villforth2014} and \cite{Villforth2017} find no difference between the morphologies of AGN hosts and control galaxies. Though this is far from a complete summary of the literature, thus far observational evidence is inconclusive with regards to the question of a merger-AGN connection. The work presented here seeks to extend these analyses to higher luminosities and larger redshift ranges than previously attempted in a single coherent study (as opposed to a compilation of various surveys and analyses).

In the local universe it is generally fairly easy to identify recent mergers by eye: low surface brightness features such as tidal tails are relatively easily picked out and mergers can be inferred from this highly non-equilibrium state \citep{Vorontsov1959, Cisternas2011, Pawlik2016}. In the high redshift universe it is far more challenging. Faint features may be of too low surface brightness to reliably detect, even soon after the merger, and galaxies are often poorly resolved. In addition, when looking for connections between AGN and galaxies, light from the bright point source is inevitably spread across the plane of the galaxy with some Point Spread Function (PSF) which can drown out galaxy features, especially those of low surface brightness. High resolution is thus required for morphological study in the high redshift universe, making the Advanced Camera for Surveys on the HST ideal. The COSMOS field is a $\sim$2 deg$^{2}$ region of sky utilising many telescopes for multi-wavelength coverage \citep{Scoville2007b}, including HST imaging with a point source limiting magnitude of 27.2. This is deep enough to detect even faint objects. The HST has a complex and hard to model PSF. Studies in the past have dealt with this by use of modelling programs such as TinyTim \citep{Krist2011} or empirically determining the PSF by averaging the distribution of light from stars \citep{Villforth2014}. The latter method was used in this work. 

Looking to earlier epochs means looking toward the peak of AGN activity (z$\sim$2-3), where the most rapid growth of SMBHs occurred \citep{Ueda2003, Aird2010}. It has been argued that only the most luminous AGN, with the highest accretion rates and therefore the most rapid loss of angular momentum, need to be fuelled by mergers \citep{Sanders1988, DiMatteo2005, Hopkins2008}. This makes z$\sim$2 a natural place to search for an AGN-merger connection, since z$\sim$2 also coincides with the predicted peak of galaxy mergers in hierarchical galaxy evolution \citep{Kauffmann2000, Volonteri2002, Conselice2003b}. Because of the difficulties inherent in detecting mergers at such high redshift many studies thus far have focussed on either low luminosity sources \citep[e.g.][]{Schawinski2011, Kocevski2012, Villforth2014, Villforth2017} or low redshift AGN \citep{Schawinski2010}. In this study we will analyse COSMOS HST F814W band data, comparing a sample of AGN host morphologies to control galaxies between z=0.5-2.2. It is not enough to show that AGN do or do not live in merging galaxies; the crucial comparison is to an inactive sample of galaxies matched in stellar mass, redshift and star forming properties to determine whether the merger rate is enhanced in active galaxies relative to "normal" galaxies. One aim of this study is to see if triggering mechanisms changed over cosmic time, both during and outside the peak of mergers/AGN activity. Another principal aim was to probe the triggering mechanisms of the highest luminosity AGN to see if there was some threshold accretion rate which required more catastrophic events such as galaxy mergers. 

The sample of AGN is introduced in section \ref{section:Sample}; the methodology in section \ref{sec:methods}; the results are presented in section \ref{sec:Results} and discussed in section \ref{sec:Disc} before conclusions are presented in section \ref{sec:Conclusions}. AB magnitudes are used throughout and, where needed, a $\Lambda$CDM cosmology with h$_{0}$ = 0.7, $\Omega$$_{m}$ = 0.3, $\Omega$$_{\Lambda}$ = 0.7 is assumed. It is worth noting that with this cosmology the angular scale is approximately constant at $\sim$8.5 kpc/" between z=1-2.2, dropping off sharply to $\sim$6.2 kpc/" at z=0.5. The fraction of galaxies in this work with z<1 is small enough to state here that working with pixel scales throughout is approximately equivalent to working with physical scales and will not affect the conclusions.

\section{Sample and Data}
\label{section:Sample}

The COSMOS survey contains multi-wavelength coverage over a region of sky $\sim$2 deg$^{2}$ \citep{Scoville2007b}. This includes X-ray data from XMM-Newton at a depth of $\sim$50Ks \citep{Hasinger2007, Cappelluti2009}, allowing for identification of a sample of AGN using X-rays, and HST F814W data for optical followup. X-rays are used for identification of AGN because it has been argued that they provide the cleanest AGN sample compared to other selection methods (e.g. optical, IR, radio), with very few non-AGN sources with L\textsubscript{X} > 10\textsuperscript{42} erg s\textsuperscript{-1} \citep{Ranalli2003}. Only Compton thick sources \citep[N\textsubscript{H} > 10\textsuperscript{24} cm\textsuperscript{-2};][]{Risaliti1999} may obscure X-ray emission from AGN.

HST imaging in the F814W band has a point source limiting AB magnitude of 27.2 at 10$\sigma$ and a surface brightness limiting magnitude of 26.1 mags/arcsecond\textsuperscript{2} at 3$\sigma$ (calculated from image noise), otherwise stated as a 5$\sigma$ depth in a 3" aperture of 25.3 \citep{Capak2007}. Point source depths drop of with (1+z)\textsuperscript{2}, while extended sources drop off with (1+z)\textsuperscript{4}. This high sensitivity makes the F814W HST filter ideal for identifying low surface brightness features such as tidal tails, hence was used in this investigation in the search for evidence of mergers. HST images in COSMOS have been corrected for sky background and instrumental noise, and drizzled to a resolution of 0.03 "/pixel in post-processing \citep[see][for details]{Koekemoer2007, Massey2010}. From the 545 luminous X-ray detected type 1 AGN in the COSMOS field in \cite{Lusso2009}, 106 were selected for analysis here over a redshift range z=0.5-2.2. A wide range of redshifts were selected in order to investigate if triggering mechanisms varied with time. We stipulated the need for pre-existing SMBH mass measurements, as required for development of the control sample (section \ref{subsec:Controls}). 

\begin{figure}
	\includegraphics[width=\columnwidth]{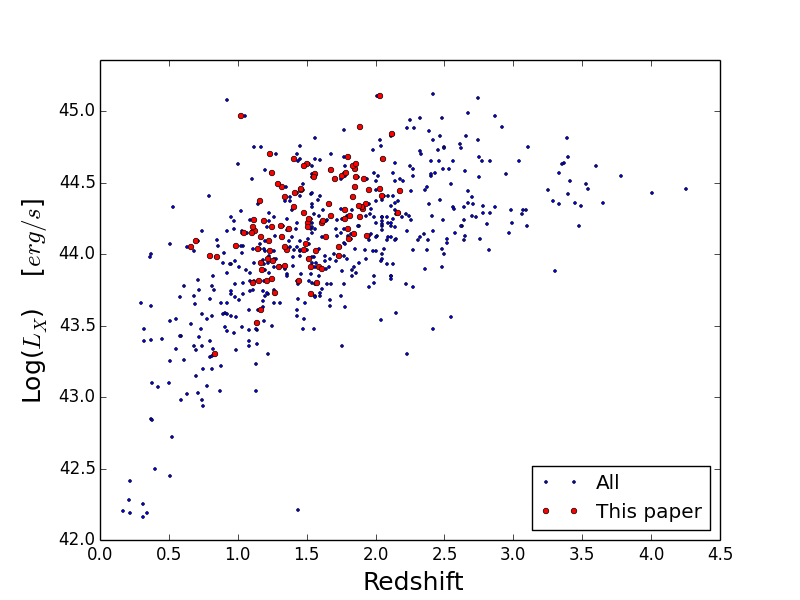}
    \caption{Larger red points represent the X-ray luminosities and redshifts of AGN in this sample; blue points show the total sample they were selected from. X-ray luminosities and redshifts are taken from \citet{Lusso2009}.}
    \label{fig:Xray_Z}
\end{figure}

The 106 AGN selected are spectroscopically confirmed to lie between z=0.5-2.2 \citep[][and references therein]{Lusso2009} and have estimates of the SMBH masses from line widths of the Mg II lines \citep{Peterson2004}. Few AGN beyond z=2.2 have black hole mass measurements due to the difficulty of obtaining high quality spectra, hence we limit our sample to AGN with z<2.2. Figure \ref{fig:Xray_Z} shows the complete \citet{Lusso2009} sample of AGN with those selected for this study in red. Control galaxy selection is discussed in section \ref{subsec:Controls}.

\graphicspath{ {/users/timhewlett/Documents/first_year_plots/} }

\section{Methodology}
\label{sec:methods} 

For the sample we now aim to derive galaxy properties, such as the dominance of disks or bulges, by fitting to the light profiles; building a sample of control galaxies with similar properties to AGN hosts; and subtracting models of the light profiles, using the residual flux and visual classification to measure how disturbed the galaxies are. The residual flux method is qualitatively described in section \ref{ResidualFlux} below, but is comprehensively described in section \ref{subsec:ResidualFlux}, along with potential biases associated with it in section \ref{subsec:Bias}. Use of \textsc{galfit} to construct 2D galaxy light profiles is described below in section \ref{galfit}; the selection and alteration of control galaxies to ensure they are equivalent to AGN hosts is described in section \ref{subsec:Controls}. Details of human visual classification are shown in section \ref{subsec:VisualClassification}.

\subsection{2D Decomposition}
\label{galfit}

In this section we describe the fitting procedure within \textsc{galfit}. We derive galaxy properties such as Sersic indices and effective radii, magnitudes of galaxies and point sources. This provides a diagnostic of the evolutionary state of galaxies; for instance, a galaxy with Sersic index n=4 is most likely an evolved elliptical, n=1 corresponds to a disk. Subtracting best fitting models from images reveals previously hidden structure, and measurement of the residual flux provides a quantitative measure of disturbance (sections \ref{ResidualFlux} and \ref{subsec:ResidualFlux}). However, in order to accomplish any of this effectively a good PSF must first be constructed.

We constructed a PSF from field stars. 13 stars with a range of apparent magnitudes (m\textsubscript{star} $\sim$ 20 - 23) were initially selected for this by identifying stars in the SDSS field which overlapped with the COSMOS field; 5 were rejected due to possible blended sources, leaving 8 stars. These were combined and normalised to a total flux of 1 within a circle of radius 50 pixels, slightly larger than a typical galaxy radius (with half light radii typically <10 pixels). This avoids artificial structure in the residual images due to over-subtraction in an arbitrarily small region and resulting in a smooth subtraction of the point source, despite only small fractions (<1\%) of point source light lying beyond a radius of 30 pixels. In the outer wings of the PSF the signal to noise ratio is low, around 1.5, with most flux concentrated in the centre. Application of this PSF to each of the stars used to construct it demonstrated reliable subtraction, leaving residuals consistent with noise, and so 8 was deemed a sufficient number of stars to accurately build the PSF. It was not clear that more stars would have improved the accuracy of the PSF, given its efficacy in tests.

Models of the light profiles of galaxies were made with the two dimensional (2D) fitting program \textsc{galfit} \citep{Peng2002, Peng2009} and subtracted from the images. In the models galaxy surface brightness varies smoothly and was constrained to be symmetric about the centre. \textsc{galfit} operates through a $\chi$\textsuperscript{2} minimisation algorithm to find the best fit parameters for point sources, Sersic profiles, spiral arms and more complex components as necessary \citep{Peng2002, Peng2009}. Models in this study were kept as simple as was required to accurately recreate light profiles and features such as spiral arms or bars were not fit; models consisted only of point sources and Sersic profiles. 

It was already known that these galaxies contained a point source (the AGN) and so the initial model in each case was a point source only. Figure \ref{fig:AddStar} demonstrates how removing the point source can reveal a previously hidden underlying galaxy. A Sersic profile was introduced in addition where necessary, with models kept as simple as required to accurately recreate the light profile. Companion galaxies (arguably foreground/background galaxies, not actively merging with the primary target) were masked out to improve fits. Models were discriminated between based on the value of the reduced $\chi$\textsuperscript{2} and the physical plausibility of parameters. Specifically, since especially large Sersic indices/small half light radii are degenerate with point sources, and small Sersic Indices/large half light radii tend toward fitting noise, it was stipulated that only Sersic models with 0.6 < R\textsubscript{e} [pixels] < 70 ($\sim$0.15-18kpc) and 0.3 < n < 10 would be accepted, and only then if the reduced $\chi$\textsuperscript{2} was improved relative to a point source only model. If model parameters failed to converge the fitting was run twice more with the Sersic index held fixed at n=4 (classical bulge) and n=1 (exponential disk). Fixing values reduces the number of parameters to be fit in \textsc{galfit} and so can aid in the construction of models. If convergence was still not achieved, or the reduced $\chi$\textsuperscript{2} was not improved, the point source only model was selected. The high luminosity of the AGN in this sample makes fitting in \textsc{galfit} especially difficult, and parameters often had to be fixed for fits to work. When fixed, there may be a bias toward fitting exponential profiles (n=1). Since compact bulges are degenerate with point sources \citep[e.g.][]{Bruce2015}, exponential profiles may more smoothly capture the outer regions of a galaxy while the point sources recreate the central regions. Getting the combination of the correct luminosity and radial extent of a classical bulge fit in combination with a bright point source is extremely challenging. We note that fixing parameters introduces some subjectivity, and this makes direct comparison to the fits of other lower luminosity, lower redshift samples problematic \citep[e.g.][]{Gabor2009}. For $\sim$50\% of galaxies a point source alone was sufficient to model the light profile, since the underlying galaxy was too faint to be detected/modelled. The other $\sim$50\% were fit by a point source + Sersic profile.

\subsection{Control Galaxies}
\label{subsec:Controls}

Selecting a group of control galaxies similar to the AGN hosts was imperative since the enhancement of merger rates in AGN hosts relative to inactive galaxies was the quantity of interest, rather than the absolute value of merger rates. Similar galaxies may have comparable merger rates but no AGN, which would suggest AGN are triggered by something other than mergers, and any presence of mergers is coincidental. AGN accretion rates are expected to be a function of stellar mass and redshift \citep{Netzer2007}, as are major merger rates \citep{Cedric1994, Fakhouri2010}, so galaxies were matched in stellar mass and redshift. By comparing to a carefully matched control group the relative levels of disturbance for different AGN properties (e.g. luminosity) could be compared.

Control galaxies were selected from the COSMOS Photometry Catalogue January 2006 \citep{Mobasher2007}. Masses and redshifts in this catalogue were determined from photometry (M/L$_{V}$ $\propto$ B-V; mass to light ratios are proportional to colour, so masses can be estimated from measurements of V band luminosity and B-V colours). There is remarkable agreement between spectroscopic and photometric redshifts, with an rms scatter of $\Delta$(z) = (z\textsubscript{phot} - z\textsubscript{spec})/(1+z\textsubscript{spec}) = 0.031 between them \citep[see][for details]{Mobasher2007}.

AGN host galaxy masses were estimated from the SMBH mass using the M-M\textsubscript{$bulge$} relationship from \citet{McConnell2012} (equation \ref{eq:M-sigma}). This was done because many AGN hosts lacked stellar mass estimates and obtaining stellar masses is especially difficult when the light is so dominated by AGN emission. Equation \ref{eq:M-sigma} provides a consistent mass estimate for all AGN hosts.

\begin{equation}
   \frac{\log(M_{\mathrm{bh}})}{M_{\odot}}=(8.46\pm{0.08})+(1.05\pm{0.11})\log\Bigg(\frac{M_{\mathrm{bulge}}}{10^{11} M_{\odot}}\Bigg)
	\label{eq:M-sigma}
\end{equation}

\begin{figure}
	\includegraphics[width=\columnwidth]{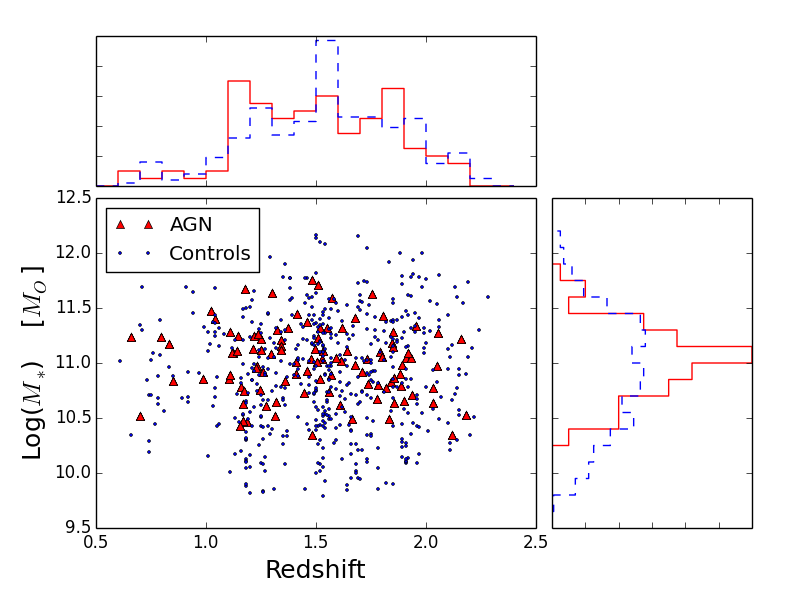} 
    \caption{Estimated stellar masses of AGN hosts (red triangles) and the matched control galaxies (blue circles). Projected histograms show the distribution of AGN and control galaxies in stellar mass and redshift; blue dashed histograms correspond to control galaxies, red solid histograms to AGN hosts.}
    \label{fig:MassZ}
\end{figure}

A Gaussian distribution around each AGN host galaxy mass was created, with width defined by scatter in equation \ref{eq:M-sigma}, and five mass values were picked at random from this gaussian. These values were matched as closely as possible to masses and redshifts in the COSMOS Photometry Catalogue leading to 5 matches per AGN (4 in one case due to a lack of 5 matching galaxies) and an initial control sample of 529 galaxies.  Figure \ref{fig:MassZ} demonstrates the outcomes of the matching procedure. To avoid AGN contamination in the control group the 43/529 that were X-ray detected were removed from further analysis, leaving 486 control galaxies remaining in the sample with a minimum of 3 controls per AGN, the majority of AGN still matched to 5 control galaxies. The X-ray detected galaxies equally occupied the full parameter space, with no trend with mass or redshift, so no bias is introduced from some AGN hosts matching to more control galaxies than others. The maximum separation in redshift was $\Delta$z = 0.3 for rarer high mass galaxies, with the majority matched within $\Delta$z = 0.1. There is a larger range in mass due to the scatter in equation \ref{eq:M-sigma}, with the median mass range about the assumed host mass $\Delta$log(M/M\textsubscript{$\odot$}) $\sim$ 0.4 (with all $\Delta$log(M/M\textsubscript{$\odot$}) < 0.7), but since the true mass of the AGN hosts is unknown, merely estimated from equation \ref{eq:M-sigma}, and there is large uncertainty in the stellar masses from the B-V colour, the intention was to compare to a range of control galaxies, each of which may represent the true form of the AGN host. As long as the average masses are similar, and the M-M\textsubscript{$bulge$} relation approximately holds to z$\sim$2 \citep{Robertson2006b, Lauer2007, Shankar2009, Shen2015}, the control galaxies should, on average, represent the AGN hosts. We appreciate this is a mildly controversial assumption; however, current evidence suggests relations between the mass of the SMBH and the galaxy it inhabits must be slowly evolving, if at all \citep{Robertson2006b, Shen2015}, and so uncertainties in control galaxy mass will dominate. Projected histograms in figure \ref{fig:MassZ} show the distribution of redshifts and (in the case of AGN hosts, assumed) masses for AGN and matched control galaxies. As expected, the match in redshift is extremely tight while there are longer tails in the control galaxy masses due to selecting masses randomly from a gaussian distribution around the expected value of the AGN host mass. There is a slight bias in the matching because lower mass galaxies are more numerous, so in matching to the first available galaxy within some bounds (e.g. $\pm$log(0.1)) it will more likely be of slightly lower mass than the AGN host. The effect is small however, with mean masses log(M$_{\text{control}}$) = 10.9M$_{\odot}$ and log(M$_{\text{AGNhost}}$) = 11.0M$_{\odot}$.

The F814W band samples the NUV light at z$\sim$2, and optical ($\lambda$$\sim$5000$\AA$) at z$\sim$0.5, so will predominantly image star forming regions. Star formation is clumpy at high z \citep{Cowie1995, VanDenBergh1996} so this may produce patchier light profiles at high redshift, contributing to the residual flux without any morphological disturbance. The dominance of AGN light in the spectra makes controlling for star forming properties extremely difficult. AGN hosts have a range of star formation rates \citep[e.g.][]{Mullaney2012}. Rest frame U-V and U-B colours reveal the same is likely true for the control galaxies here, with colours representative of the underlying distribution, and so it is likely that the star forming properties of the control galaxies match reasonably well to the AGN hosts. It is noted that star formation rates are not directly matched between AGN hosts and control galaxies, but that the control galaxies are representative of galaxies of similar mass.

\begin{figure*}
	\centering
	\begin{minipage}[t]{\dimexpr.9\textwidth-1em}
	\centering
	\includegraphics[width=\textwidth]{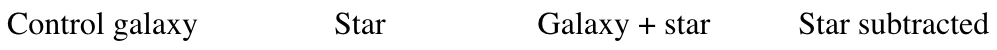}
	\includegraphics[width=0.2\textwidth]{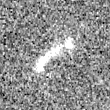}\hfill
	\includegraphics[width=0.2\textwidth]{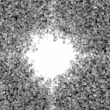}\hfill
	\includegraphics[width=0.2\textwidth]{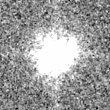}\hfill
	\includegraphics[width=0.2\textwidth]{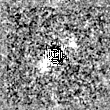}\hfill
	
	\includegraphics[width=0.2\textwidth]{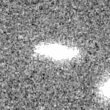}\hfill
	\includegraphics[width=0.2\textwidth]{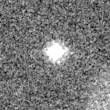}\hfill
	\includegraphics[width=0.2\textwidth]{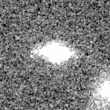}\hfill
	\includegraphics[width=0.2\textwidth]{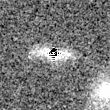}\hfill
	
	\includegraphics[width=0.2\textwidth]{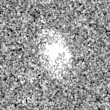}\hfill
	\includegraphics[width=0.2\textwidth]{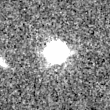}\hfill
	\includegraphics[width=0.2\textwidth]{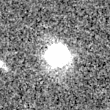}\hfill
	\includegraphics[width=0.2\textwidth]{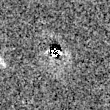}\hfill
	
	\includegraphics[width=0.2\textwidth]{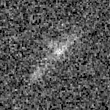}\hfill
	\includegraphics[width=0.2\textwidth]{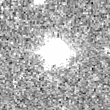}\hfill
	\includegraphics[width=0.2\textwidth]{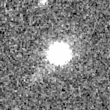}\hfill
	\includegraphics[width=0.2\textwidth]{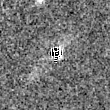}\hfill
	
	\includegraphics[width=0.2\textwidth]{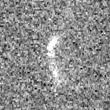}\hfill
	\includegraphics[width=0.2\textwidth]{Star11}\hfill
	\includegraphics[width=0.2\textwidth]{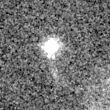}\hfill
	\includegraphics[width=0.2\textwidth]{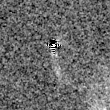}\hfill
	
	\includegraphics[width=0.2\textwidth]{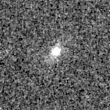}\hfill
	\includegraphics[width=0.2\textwidth]{Star4}\hfill
	\includegraphics[width=0.2\textwidth]{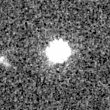}\hfill
	\includegraphics[width=0.2\textwidth]{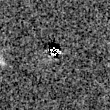}\hfill
	
	\includegraphics[width=0.2\textwidth]{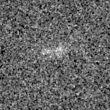}\hfill
	\includegraphics[width=0.2\textwidth]{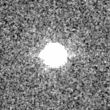}	\hfill
	\includegraphics[width=0.2\textwidth]{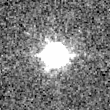}\hfill
	\includegraphics[width=0.2\textwidth]{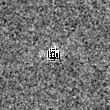}\hfill
	
    \caption{$\sim$3" cutouts of control galaxy (left) plus star (centre left) equals final image (centre right) for seven example galaxies. Far right panels show the stars being re-subtracted from these images, using \textsc{galfit}, to recover the host galaxy profile. From top to bottom: ID=2381537, z=1.56; ID=3024848, z=1.34; ID=2747981, z=1.24; ID=2479182, z=1.56; ID=688479, z=1.55; ID=3038344, z=2.06; ID=427649, z=1.53.}
    \label{fig:AddStar}
\end{minipage}
\end{figure*}

Many images of AGN + hosts are dominated by light from the point source, making distinguishing features by eye difficult. To make control galaxies actually resemble AGN hosts, 26 stars were identified from the SDSS, selected in COSMOS, and matched as closely as possible in apparent magnitude to AGN and added to the control galaxies as AGN-like point sources. Figure \ref{fig:AddStar} shows control galaxies before and after adding stars; the structure of the underlying galaxy is largely obscured by even a relatively dim point source, so it was imperative to add realistic point sources to the control galaxies before any further analysis. Figure \ref{fig:AddStar} also demonstrates the efficacy of removing the point source in \textsc{galfit}, successfully recovering much of the underlying galaxy structure, albeit with less certainty in the central regions. There were no adequately faint stars in the field for the faintest AGN. In this case, the faintest star available was used as a surrogate. Potential bias introduced by this is discussed in section \ref{subsec:Bias}.

\graphicspath{ {/users/timhewlett/Documents/first_year_plots/ControlGalaxiesAndStars/} }

\graphicspath{ {/users/timhewlett/Documents/first_year_plots/} }

Some ($\sim$40, 8\%) control galaxies were undetected at F814W. Their reported magnitudes in the catalogue were in the dim tail of the whole distribution of galaxies used here, $\sim$25th magnitude, so if slightly extended would be hard to pick out visually. Since stellar masses were required for selection, objects were selected from the COSMOS Photometry Catalogue 2006 \cite{Mobasher2007} instead of the ACS catalogue, and resultantly there are some cases where the galaxy is missing in ACS. We wished to avoid biasing the results, since it is likely that in some AGN images the host is also undetected in F814W with only the AGN itself seen, so a point source was added to empty sky in these cases. This allows testing of the reliability of host galaxy reconstruction in \textsc{galfit}, since if more than a point source is recovered where only a point source was added this provides an estimate of the number of false positive galaxy detections. $\sim$1/4 of these had false galaxy detections with a mean magnitude m\textsubscript{Gal}=24.4 in F814W (far fainter than the mean point source magnitude for these galaxies, m\textsubscript{PS}=21.6), and radii of 2 pixels (just over the threshold radius for the model to be accepted). This implies \textsc{galfit} is ascribing some point source light to a non-existent galaxy, hence the dim and compact galaxy. Recall that $\sim$50\% of all images, AGN and controls alike, were best fit by point sources alone, implying galaxies were often dim compared to the central point source.

\subsection{Residual Flux}
\label{ResidualFlux}

The residual method utilised here is conceptually simple. Models of the light profiles of galaxies were made with the two dimensional (2D) fitting program \textsc{galfit} \citep{Peng2002, Peng2009} and subtracted from the images. In the models galaxy surface brightness varies smoothly and was constrained to be symmetric about the centre; this means that any flux remaining in the residuals of an image will be due to the galaxy itself having disturbed or asymmetric light profiles, which may be indicative of morphological disruption. Regions of each galaxy may be over- or under-subtracted due to galaxy interactions. Therefore, summing the modulus of the residual flux on all pixels in some region around the galaxy (chosen to be a circle with radius 30 pixels, see appendix) following model-subtraction provides a direct numerical measure of disturbance in a galaxy. 

One motivation for developing the residual flux method was to create a more continuous measure of disturbance than easily achieved by visual inspection (although see \citet{Mechtley2016} for a creative continuous method of visual classification of disturbance), so it is not expected to yield identical results to visual classification (section \ref{subsec:VisualClassification}). Correlation between the methods, however, is required for credibility in the residual flux method. The method broadly captures morphological disturbance compared to visual classification, with some scatter, with a galaxy with residual flux greater than 2 having a >50\% chance of being classified as "merger" or "disturbed" (see section \ref{subsec:VisualClassification}), despite these constituting <20\% of the sample. Detailed description of the development, comparison to visual classification, and optimisation of the residual flux method is deferred to the appendix, \ref{subsec:ResidualFlux}, with possible sources of bias discussed in \ref{subsec:Bias}.

\subsection{Visual Classification}
\label{subsec:VisualClassification}

\begin{figure*}
	\centering
	\begin{minipage}[t]{\dimexpr.9\textwidth-1em}
	\centering
	\includegraphics[width=\textwidth]{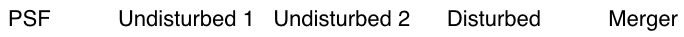}
	\end{minipage}
	\includegraphics[width=0.18\textwidth]{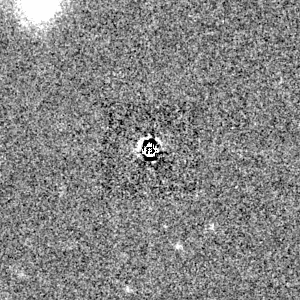} \space\space
	\includegraphics[width=0.18\textwidth]{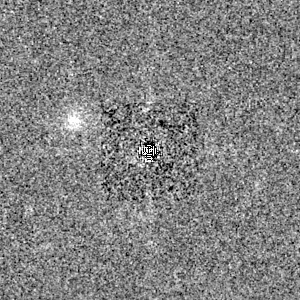} \space\space
	\includegraphics[width=0.18\textwidth]{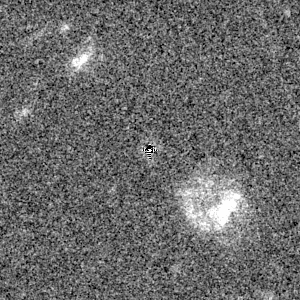} \space\space
	\includegraphics[width=0.18\textwidth]{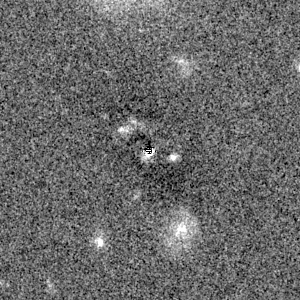} \space\space
	\includegraphics[width=0.18\textwidth]{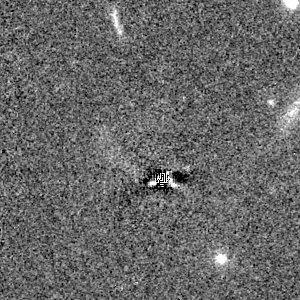} \space\space
	
	\includegraphics[width=0.18\textwidth]{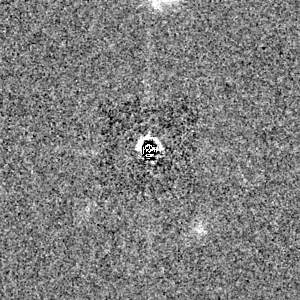} \space\space
	\includegraphics[width=0.18\textwidth]{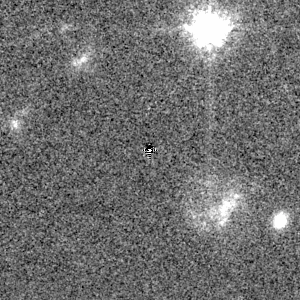} \space\space
	\includegraphics[width=0.18\textwidth]{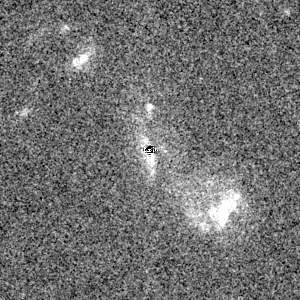} \space\space
	\includegraphics[width=0.18\textwidth]{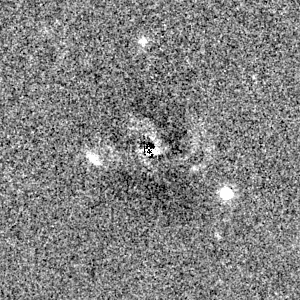} \space\space
	\includegraphics[width=0.18\textwidth]{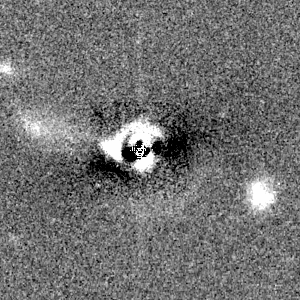} \space\space
	
	\includegraphics[width=0.18\textwidth]{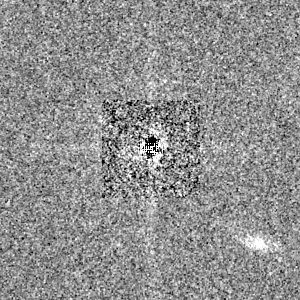} \space\space
	\includegraphics[width=0.18\textwidth]{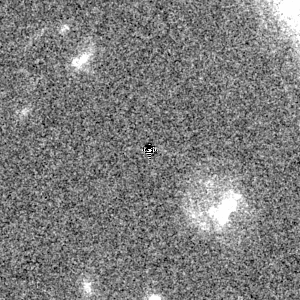} \space\space
	\includegraphics[width=0.18\textwidth]{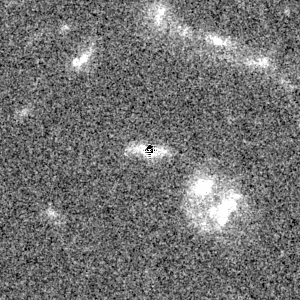} \space\space
	\includegraphics[width=0.18\textwidth]{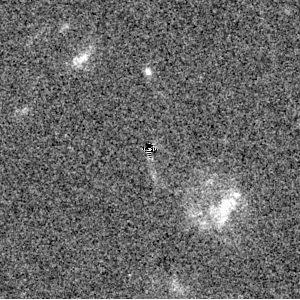} \space\space
	\includegraphics[width=0.18\textwidth]{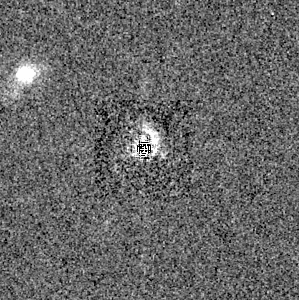} \space\space
	
	\includegraphics[width=0.18\textwidth]{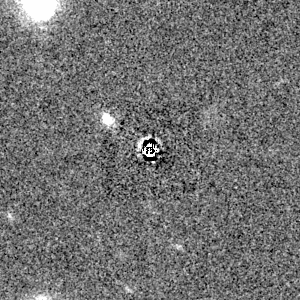} \space\space
	\includegraphics[width=0.18\textwidth]{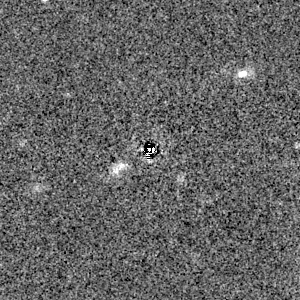} \space\space
	\includegraphics[width=0.18\textwidth]{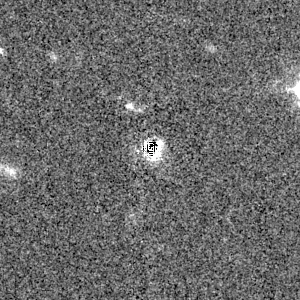} \space\space
	\includegraphics[width=0.18\textwidth]{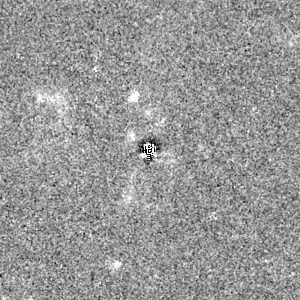} \space\space
	\includegraphics[width=0.18\textwidth]{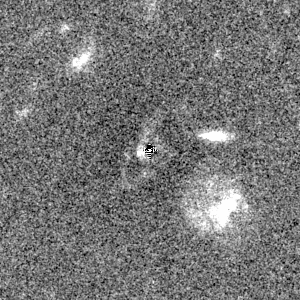} \space\space
	
	\includegraphics[width=0.18\textwidth]{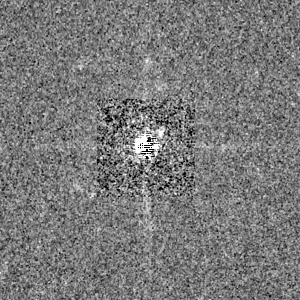} \space\space
	\includegraphics[width=0.18\textwidth]{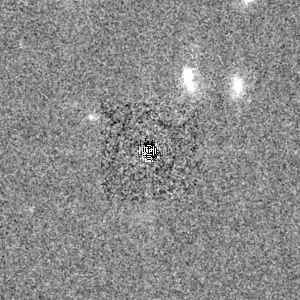} \space\space
	\includegraphics[width=0.18\textwidth]{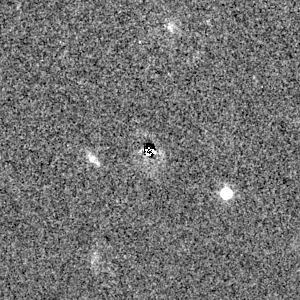} \space\space
	\includegraphics[width=0.18\textwidth]{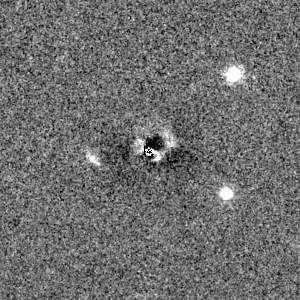} \space\space
	\includegraphics[width=0.18\textwidth]{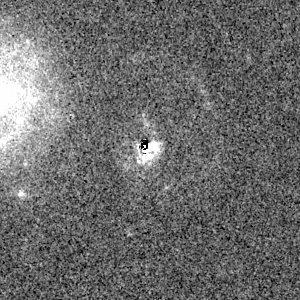} \space\space
	
	\includegraphics[width=0.18\textwidth]{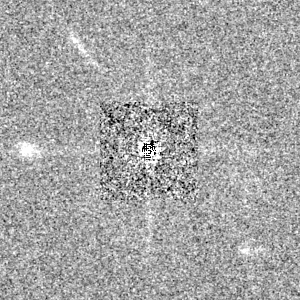} \space\space
	\includegraphics[width=0.18\textwidth]{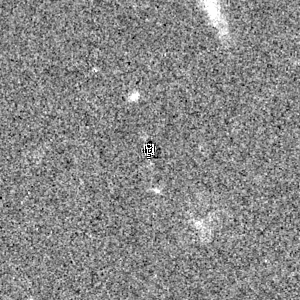} \space\space
	\includegraphics[width=0.18\textwidth]{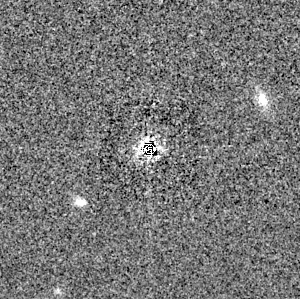} \space\space
	\includegraphics[width=0.18\textwidth]{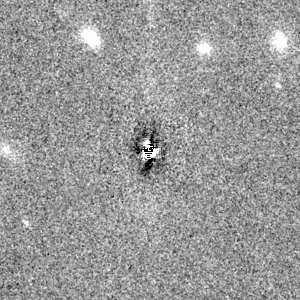} \space\space
	\includegraphics[width=0.18\textwidth]{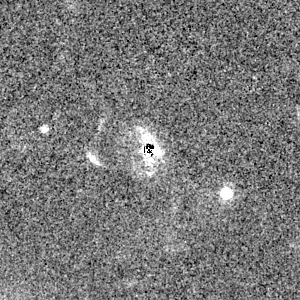} \space\space
	
    \caption{Example classifications of different galaxies. Columns from left to right: PSF dominated; undisturbed without residuals; undisturbed with residuals; disturbed; merger. Although subjective, the merger is distinguished in particular from its extended tidal features.}
    \label{fig:ClassificationExamples}
\end{figure*}

Three experts classified the galaxies into five main groups both for comparison to the residual flux method and for independent science analysis. The five main categories of classification (see figure \ref{fig:ClassificationExamples} for example classifications) are: 

\hfill

i) PSF dominated - for cases where the image is dominated by poor PSF subtraction: for instance, large symmetric diffraction spikes.

ii) Undisturbed without residuals and iii) undisturbed with residuals - classes intended to discriminate between galaxies that the residual flux method might pick out as merger candidates, but humans reveal to have highly symmetric residuals due, for example, to marginal galaxy detections.

iv) Disturbed galaxies - where there is some asymmetry or otherwise disturbance in the light profile, but the source of the disturbance may be unclear, for example with asymmetric regions of over- or under-subtraction but no clear multiple cores. This may correspond to the late stages of a merger, more minor mergers, or possibly in some cases to clumpy disks.

v) Clear mergers - a category for galaxies deemed to be clear mergers with multiple cores, tidal tails, or otherwise excessive asymmetry.

 \hfill

Classifiers were asked to put each of the 592 model-subtracted galaxies into one of the five classes, with additional optional columns for features such as spiral arms or foreground objects. Classifiers were in broad agreement with one another: when one classifier categorises a galaxy as a merger $\sim$85\% of the time the other two classifiers have classified it as disturbed or a merger. 

How does the detectability of merger features vary with redshift? \cite{Kaviraj2013b} analyse merger remnants from a cosmological hydrodynamical simulation \citep{Peirani2010} and find that at redshifts of $\sim$1.25, only mergers with mass ratios of 1:5 are detectable with HST imaging at a depth of 26 mag arcsec\textsuperscript{-2}. At z$\sim$3 this detectability reduces to real train wrecks, with mass ratios of 1:2. This suggests that those identified as mergers here are likely to be major mergers, and at higher redshifts the fraction of galaxies identified as mergers will drop off (unless balanced by an intrinsic increase in the fraction undergoing major mergers). Full constraints on the detectability of merger features through cosmic time from cosmological simulations is beyond the scope of this work and is deferred to a future publication. For this study it suffices for the control galaxies and AGN hosts to be comparable across the redshift range, giving rise to similar biases, allowing like-for-like comparisons to be made.

\section{Results}
\label{sec:Results}

\begin{figure}
	\includegraphics[width=\columnwidth]{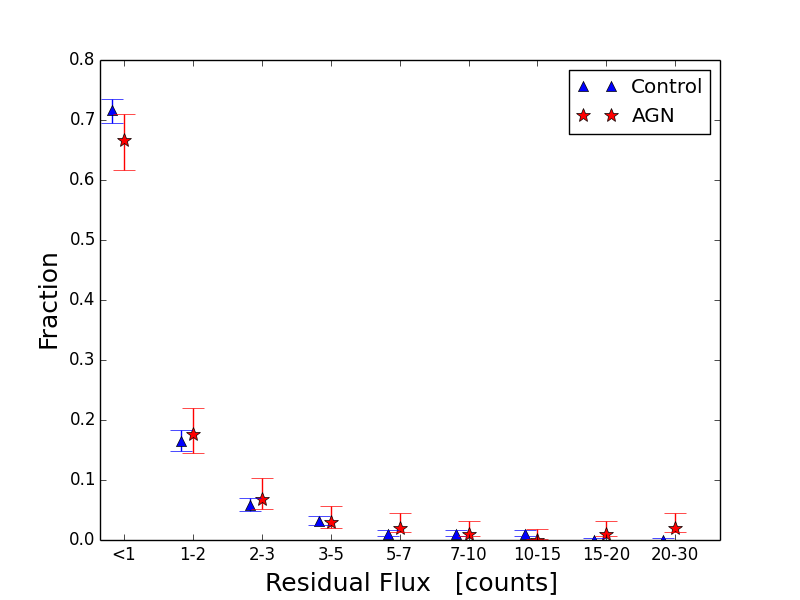}
    \caption{Fraction of AGN and control galaxies in bins of residual flux. Blue triangles are control galaxies and red stars represent AGN. 1 sigma confidence levels are calculated according to the prescription of \citet{Cameron2011}.}
    \label{fig:ResidsHist}
\end{figure}

\begin{figure}
	\includegraphics[width=\columnwidth]{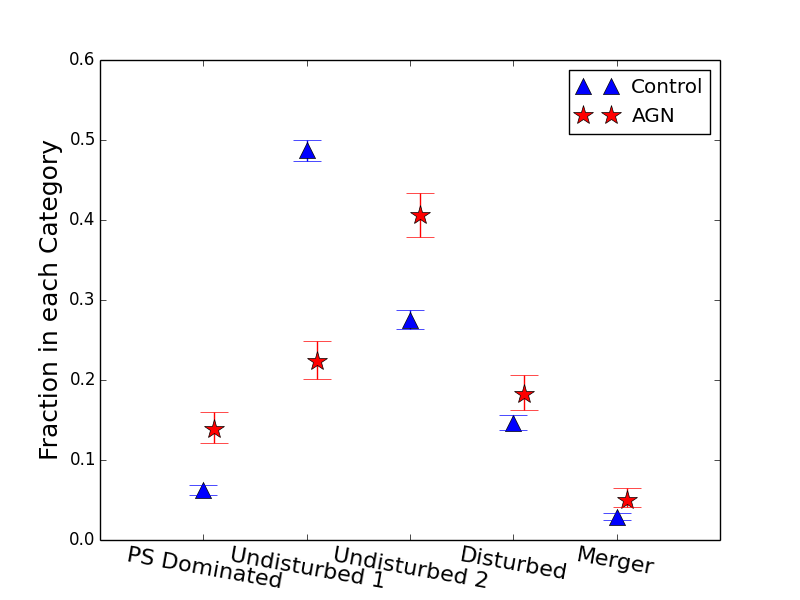}
    \caption{The total fraction of AGN and control galaxies classified by humans as mergers, disturbed, undisturbed with residuals, undisturbed without residuals or point source dominated. Blue triangles are control galaxies and red stars AGN. As in figure \ref{fig:ResidsHist} 1 sigma confidence levels are calculated according to the prescription of \citet{Cameron2011}.}
    \label{fig:ClassificationFractions}
\end{figure}

We will now present the results of the residual flux method alongside visual classifications to determine any difference between morphological disturbance of AGN hosts and control galaxies as a function of: 1) the fraction of control galaxies and AGN in different categories or residual flux bins; 2) the point source or AGN luminosities; 3) redshift. For each question we will first present evidence from the residual flux method, then from visual classification. 

The most obvious question we can ask regards statement 1: are AGN more likely to live in disturbed systems? Figure \ref{fig:ResidsHist} shows the comparison between AGN hosts and control galaxies in residual flux. By this metric AGN are no more likely to be found in disturbed galaxies than control galaxies. Similarly, figure \ref{fig:ClassificationFractions} shows no statistically significant difference between controls and AGN hosts in visual classification, with both merger and disturbed fractions equivalent for both samples. Together these results show that there are not higher numbers of AGN in merging galaxies in this sample. 

Through fitting to the light profiles in \textsc{galfit}, it was found that the majority of galaxies in this sample are best fit by disks, rather than spheroids, with a mean Sersic index of $\sim$1.15 for both AGN hosts and control galaxies. It might be expected for mergers to destroy disks, or otherwise make galaxies more bulge-dominated \citep{Toomre1972, Barnes1988}. Finding comparable disk components in AGN hosts and control galaxies suggests that these AGN do not live in especially disturbed or evolved systems. However, as described in section \ref{galfit} models were often improved by holding some parameters fixed, so the calculated mean Sersic index may be somewhat biased by this and care should be taken in interpreting the results.

\begin{figure}
	\includegraphics[width=\columnwidth]{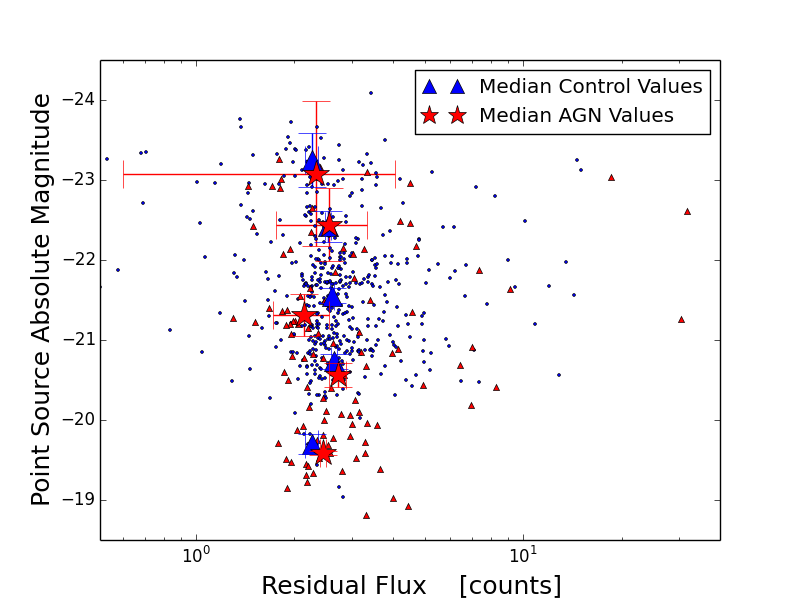}
    \caption{Residual flux of AGN + hosts and control galaxies as a function of the point source absolute magnitude, corrected for poor subtraction of point sources. Small blue circles and red triangles represent all values of control galaxy and AGN residual flux; large blue triangles and red stars are the corresponding median values of residual flux in bins of point source absolute magnitude. All residual fluxes values have had +2 added for clarity; this is merely because $\sim$15\% of points have negative values once the erroneous point source contribution has been subtracted, and these points do not appear on a logarithmic scale. The significant comparison is between the relative median values, which is unaffected by a uniform addition. Errors are as in figure \ref{fig:ResidsClassCorrelation}.}
    \label{fig:ResidsAbsMags}
\end{figure}

\begin{figure}
	\includegraphics[width=\columnwidth]{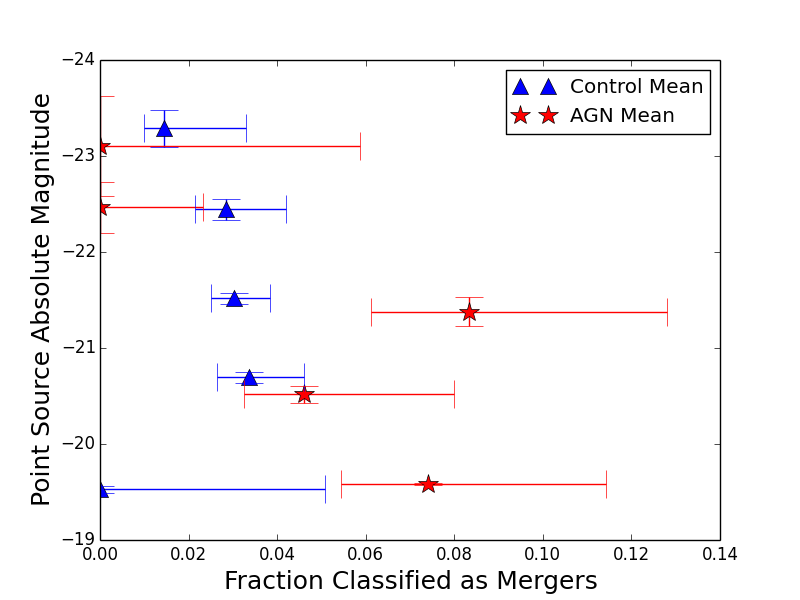}
	\includegraphics[width=\columnwidth]{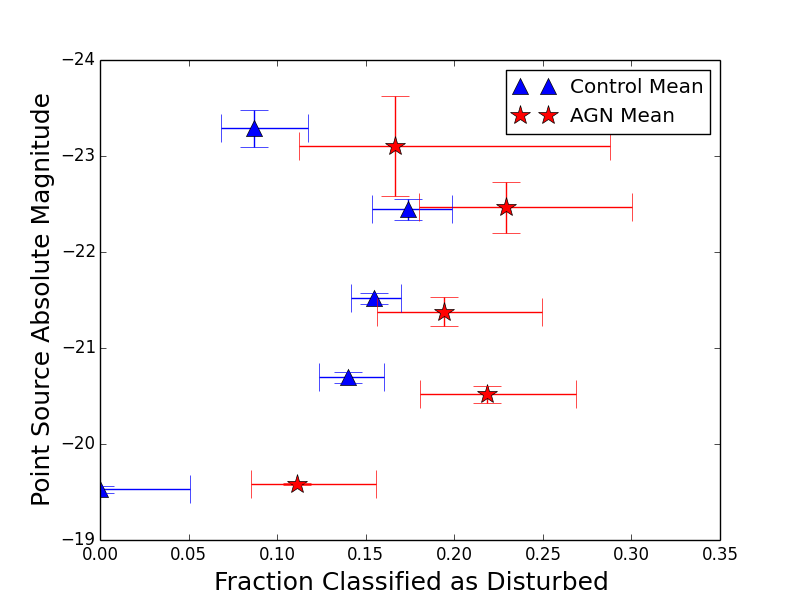}
    \caption{The fraction of galaxies classified by humans as mergers (upper panel) or disturbed (lower panel) in bins of point source absolute magnitude. 1 sigma confidence levels are calculated according to the prescription of \citet{Cameron2011}.}
    \label{fig:ClassificationMagnitude}
\end{figure}

To address point 2), the question of the impact of mergers on AGN luminosity: if the most powerful AGN must be triggered by merging galaxies then in the highest luminosity bins the AGN should be in the clearest mergers (highest residual flux) compared to the control galaxies. Figure \ref{fig:ResidsAbsMags} shows the residual flux, corrected for PSF residuals, of AGN hosts and control galaxies as a function of the absolute magnitude of the point source. Control galaxy artificial point sources have the "absolute magnitude" that the point source would need in order to have the measured apparent magnitude at its host galaxy redshift. No enhancement in residual flux is found with increasing AGN luminosity relative to the control group. Only $\sim$3\% of galaxies are found to be disturbed above 5 counts of residual flux, but this is independent of the AGN/artificial point source absolute magnitude. If mergers were triggering the most luminous AGN the expectation would be for an enhancement of residual flux in AGN hosts over their control galaxy counterparts. 

Figure \ref{fig:ClassificationMagnitude} shows the fraction of galaxies classified as mergers or as disturbed by three human classifiers in bins of point source absolute magnitude. In the highest luminosity bins there is no merger excess, while $\sim$3\% of control galaxies at all luminosities are found in merging galaxies. Both controls and AGN hosts are consistent with a constant fraction of mergers ($\sim$3\%) or disturbed ($\sim$15\%) galaxies at all point source magnitudes. No link is found between the most luminous AGN and galaxy mergers. For a comparison between residual flux and visual classification see section \ref{subsec:ResidualFlux}.

\begin{figure}
	\includegraphics[width=\columnwidth]{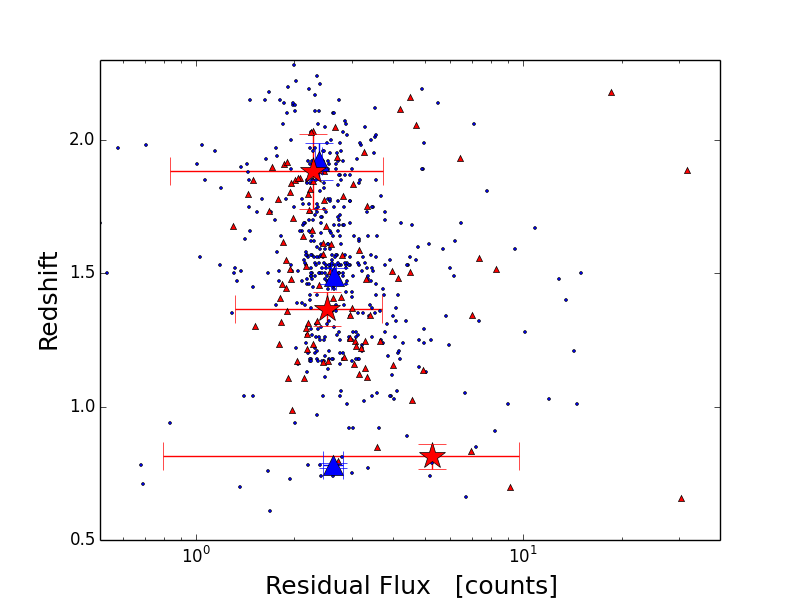}
    \caption{Relation between redshift and residual flux. As in other plots blue points are control galaxies; small red triangles are AGN; and the large blue triangles and large red stars represent the median values of controls and AGN respectively, with associated bootstrap errors for different bins of redshift. As with figure \ref{fig:ResidsAbsMags} all residual flux values have had +2 added for clarity; this is merely because $\sim$15\% of points have negative values once the erroneous point source contribution has been subtracted, and these points do not appear on a logarithmic scale. The significant comparison is between the relative median values, which is unaffected by a uniform addition. Errors are as in figure \ref{fig:ResidsClassCorrelation}.}
    \label{fig:RedshiftEvolution}
\end{figure}

\begin{figure}
	\includegraphics[width=\columnwidth]{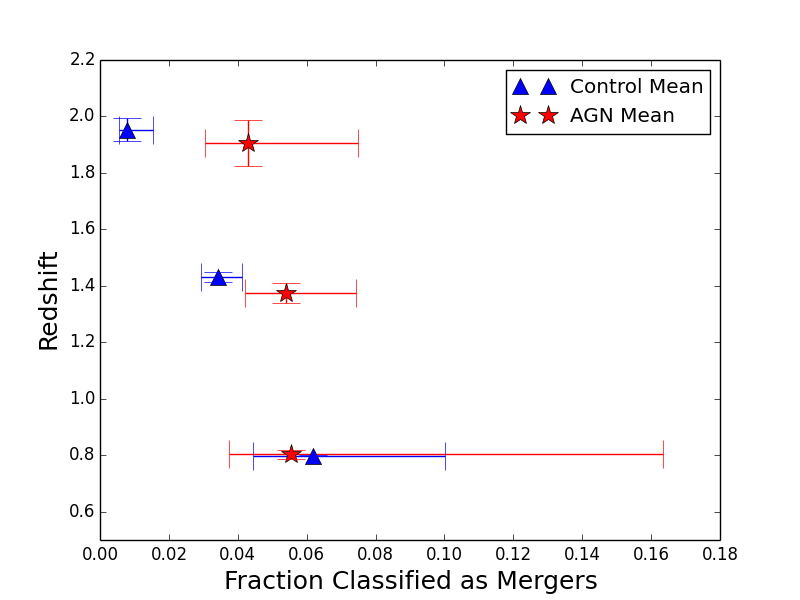}
	\includegraphics[width=\columnwidth]{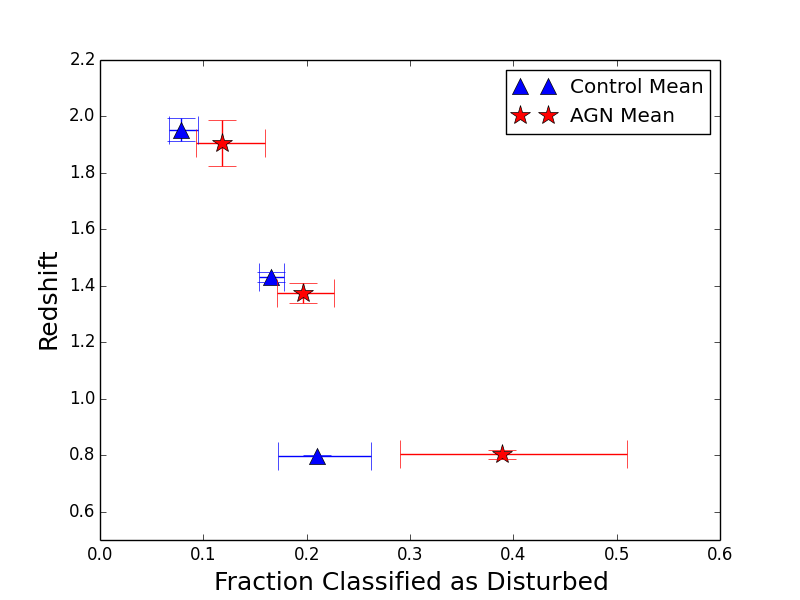}
    \caption{The fraction of galaxies classified by humans as mergers (upper panel) or disturbed (lower panel) in bins of redshift. 1 sigma confidence levels are calculated according to the prescription of \citet{Cameron2011}.}
    \label{fig:ClassificationRedshift}
\end{figure}

One of the primary aims of this investigation was to see if there was any evolution of triggering mechanisms with redshift. Figure \ref{fig:RedshiftEvolution} shows the correlation between residual flux and redshift, with similar residual flux in all redshift bins. We will now describe the enhancement of AGN in galaxies visually classified as mergers. Figure \ref{fig:ClassificationRedshift} shows the fraction of galaxies with "disturbed" or "merger" classifications in bins of redshift. A factor of $\sim$4 enhancement of AGN in galaxies visually classified as mergers relative to the control galaxies is detected at z$\sim$2 at 99\% confidence level, 2.4$\sigma$. This was determined from the binomial confidence limits as prescribed in \citet{Cameron2011}. We note the apparent evolution in the upper panel of figure \ref{fig:ClassificationRedshift}, with a smooth transition to increased merger dominance for AGN at higher redshifts, though only the highest redshift bin is statistically significant. Combining the "merger" and "disturbed" classifications, thereby improving number statistics but possibly slightly reducing the robustness of the classification, gives a factor of two enhancement of AGN in morphologically disrupted galaxies at 98.5\% confidence level. This provides evidence that AGN are more common in disturbed galaxies at high redshift, where mergers are most common \citep{Cedric1994, Fakhouri2010}, but at lower redshifts other processes are dominant. This enhancement of merger fraction at z$\sim$2 has not been found in other studies \citep{Gabor2009, Allevato2011, Cisternas2011, Schawinski2011, Kocevski2012, Treister2012, Villforth2014, Bruce2015, Villforth2017}. This is the largest study to date of the morphologies of high redshift luminous AGN. The better number statistics may allow for detection of this mild enhancement; this possibility is explored in some detail in the discussion below.

\section{Discussion}
\label{sec:Disc}

In this study 106 high luminosity AGN in COSMOS are host galaxy mass and redshift matched to 486 control galaxies to compare their morphological properties in order to attempt to answer the question: are mergers a dominant AGN triggering mechanism? \textsc{galfit} models of galaxy light profiles are made. Models are subtracted from images, and the residual flux serves as a measure of disturbance in the galaxies. Galaxies are visually classified by three experts, identifying likely mergers. Comparison to a group of mass and redshift matched control galaxies allows for analysis of the role of galaxy mergers in triggering AGN. Most galaxies are found to be best fit by exponentially declining light profiles, indicative of disk dominated morphologies \citep{Lin1987}, in contradiction to post-merger morphologies as predicted by some models \citep{Toomre1972, Barnes1988, DiMatteo2005, Cox2008}. It is found that the fractions of AGN and control galaxies are equivalent for all residual flux values. Similarly, the fractions of AGN and control galaxies in different visual classes are equivalent. In addition, no correlation is found between AGN luminosity and residual flux, and visual inspection of morphology yields no indication that more luminous AGN are hosted by merging galaxies. Residual flux values are approximately constant across the whole range of redshifts, but there is a 2.4$\sigma$ enhancement of the visually classified merger fraction of AGN hosts relative to control galaxies at z$\sim$2. This may imply a possible evolution in the dominance of merger-triggering through time. These results will now be discussed in the context of the wider literature.

First we consider the impact of selection bias on the results. The AGN studied here are X-ray selected. Are these representative of the underlying AGN population? It has been argued that merger induced AGN are more likely to be Compton-thick \citep{Cattaneo2005, Kocevski2015}, therefore selecting AGN in the X-ray biases against observing merger signatures contemporaneously with the AGN emission, which is obscured. It is important to bear this in mind when interpreting these results, but also to recall biases associated with other selection methods. Infrared (IR) emission, for instance, allows for selection of AGN by relying on the approximately power-law spectrum of AGN in the IR \citep{Stern2012}. This can, however, bias in favour of finding mergers, since mergers can induce starbursts \citep{Mihos1994, Mihos1996, Hopkins2006a} which emit powerfully in the IR, enhancing detection rates. Indeed, studies of IR selected AGN have found enhancements of AGN fractions in merging galaxies relative to controls \citep[e.g.][]{Satyapal2014}. Selecting AGN using X-rays may bias results away from finding mergers, but certain results remain true regardless: for example, that mergers are not \textit{required} to power the most luminous AGN, since we find so many luminous AGN unrelated to mergers.

What do the \textsc{galfit} models reveal about the galaxies in this sample? Some models predict that major galaxy mergers should destroy disks \citep{Toomre1972, Barnes1988, DiMatteo2005, Cox2008} and post-mergers are thought to relatively quickly evolve into elliptical galaxies \citep[unless gas-rich enough to quickly reform a disk][]{Robertson2006}. Elliptical galaxies have Sersic indices of $\sim$4 \citep{Mellier1987} while disk dominated systems have Sersic indices closer to 1 - an exponentially declining light profile \citep{Lin1987}. If the AGN hosts have recently undergone a merger, driving the nuclear activity, the Sersic indices may be expected to tend toward values indicative of bulge dominated systems. \textsc{galfit} models reveal an average Sersic index of $\sim$1.15 with similar Sersic index distributions for AGN and control galaxies, suggesting that most of the galaxies in this sample are disk dominated. As mentioned in section \ref{galfit} these models may be biased by the requirement to fit a bright point source, otherwise this provides some evidence that these are not mergers with faint merger features, or late mergers where features have faded, but instead are predominantly secularly evolving galaxies. This is in agreement with \cite{Cisternas2011}; \cite{Schawinski2011}; \cite{Kocevski2012}; \cite{Villforth2014} and \cite{Villforth2017}, extending the work to greater redshift and/or luminosity. 

Next we shall discuss the relative fractions of AGN and control galaxies in different morphological categories, and their distributions in residual flux. Figures \ref{fig:ResidsHist} and \ref{fig:ClassificationFractions} demonstrate the similarity between the AGN host and control galaxy morphologies. Similar fractions of active and inactive galaxies are found in merging or disturbed systems. Such a small fraction of AGN are found to exist in merging galaxies that the majority of black hole growth must occur via alternative processes, or the AGN must be triggered after merger features have faded. 

In addition, it is found here that even the most luminous AGN are no more likely to be found in merging galaxies (see figures \ref{fig:ResidsAbsMags} and \ref{fig:ClassificationMagnitude}), in contradiction to the conjecture that major merger triggering of AGN should be a strong function of luminosity \citep{Finn2001, Treister2012} with the most luminous AGN triggered solely by mergers. This is in opposition to some previous studies \citep[e.g.][]{Treister2012, Ellison2016} and in agreement with others \citep[e.g.][]{Cisternas2011, Allevato2011, Karouzos2014, Villforth2017}. We cannot conclusively rule out the possibility that a time delay between mergers and AGN switching on is responsible for merger features significantly fading, however the disk-like morphologies suggest major mergers have not recently occurred in the majority of the sample, making this more unlikely \citep[other studies have also come to this conclusion due to the dominance of disks in the samples, e.g.][]{Gabor2009, Cisternas2011, Schawinski2011, Kocevski2012, Villforth2014, Villforth2017}. At high redshifts, where gas densities are higher, some simulations show that these galaxies may be able to relax into disks again \citep{Robertson2006}, though since this may require specific orbital configurations, and not all the galaxies investigated here are at especially high redshift, this seems implausible to account for such a large portion of the sample here. {\citet{Treister2012} found a strong correlation between AGN luminosity and merging status, with just 4\% of AGN with bolometric luminosities of 10$^{43}$ erg s$^{-1}$ in mergers and $\sim$90\% in mergers at 10$^{46}$ erg s$^{-1}$. Whilst this appears to support the notion that the highest luminosity AGN are triggered by mergers, a range of AGN selection criteria were compiled into the study and in some cases the true AGN luminosities are uncertain. Additionally, the lack of a control group makes interpretation of these results problematic. 

Through visual analysis of 140 X-ray selected AGN with 1264 control galaxies with HST data in COSMOS, \citet{Cisternas2011} find no merger-AGN connection from z=0.3-1. Similarly to this work, they go further and demonstrate the implausibility of a time delay between mergers and the peak of SMBH accretion being responsible for this null result, since the morphologies are often disk dominated. {\citet{Allevato2011} measure the halo mass bias of different classes of AGN with redshift, concluding that the increase in bias with redshift (z<2.25) suggests that different AGN phases (classic Seyfert; quasar mode; radio loud) are purely related to the halo mass. Semi analytic models assuming major merger triggering of AGN cannot reproduce these biases for type 1 AGN, such that alternate triggering processes (secular; tidal disruptions; disk instabilities) may be more significant than major mergers for moderate luminosity AGN up to z$\sim$2.2 {\citep{Allevato2011}, a conclusion bolstered by the findings of this study. In short, this work broadly supports the findings of other observational studies, though differing AGN selection methods may obfuscate the issue. 

The final piece of evidence is the correlation between residual flux or visual classification and redshift in figures \ref{fig:RedshiftEvolution} and \ref{fig:ClassificationRedshift}. First we will discuss the residual flux result. Figure \ref{fig:RedshiftEvolution} shows similarity between AGN and control galaxies in residual flux across the redshift range. One may be concerned that merging galaxies will have similar enhancement of residual flux compared to galaxies with non-axisymmetric features, such as spiral arms and clumpy disks. Over this redshift range the F814W band measures rest frame UV through to visual; since star formation is intrinsically clumpy \citep{Bournaud2011, Trump2014} this may manifest itself as excess residuals at high redshift. We did not attempt to fit non-axisymmetric structures (section \ref{galfit}), instead choosing to compare residual flux to visual classification. Visual class and residual flux correlate well (see figure \ref{fig:ResidsClassCorrelation}).Visual classifiers had the opportunity to highlight non-axisymmetric structures for each galaxy: their exclusion did not affect the result, and further inspection of images with high residual flux reveals tidal tails and structures that do not look like clumpy disks or spiral arms, ruling this out as the probable cause of similarity (See figure \ref{fig:Resids5}).

Considering visual classification, figure \ref{fig:ClassificationRedshift} shows the fraction of AGN hosts and control galaxies classified as mergers in bins of redshift. AGN hosts are found to have a higher incidence of mergers at high redshift, with a factor of $\sim$4 enhancement of AGN in mergers at z$\sim$2 with 99\% confidence. Considering disturbed and merger classifications together as one class gives a factor of $\sim$2 enhancement with 98.5\% confidence at z$\sim$2. We consider whether a mass mismatch between the AGN hosts and control galaxies could lead to a difference in merger incidence between AGN hosts and controls. If, for instance, the M-M\textsubscript{bulge} relationship evolves with redshift this could lead to a mass mismatch, and galaxies of different masses may undergo major mergers with different frequencies. \cite{Stewart2009} and \cite{Rodriguez2015} predict the merger frequency to be a function of mass, with more massive galaxies having slightly higher merger rates. At z$\sim$2 there are approximately twice as many major mergers for galaxies with M\textsubscript{*}$\sim$10\textsuperscript{11}M$_{\text{$\bigodot$}}$ as there are for M\textsubscript{*}$\sim$10\textsuperscript{10}M$_{\text{$\bigodot$}}$ \citep{Rodriguez2015}. This suggests any mass mismatch here would have to be implausibly severe to entirely explain the observed factor of $\sim$4 enhancement. Were the mass mismatch so severe, the model magnitude distributions for AGN hosts and control galaxies would be extremely different to each other: in reality they are comparable. The M-M\textsubscript{bulge} relation would be expected to evolve slowly, if at all, to have avoided robust detection in previous studies \citep{Robertson2006b, Lauer2007, Shankar2009, Shen2015}, so it is likely that control and AGN host galaxies are similarly massive.

Next, we consider the possibility that this result has remained undetected in previous studies due to poor number statistics. Work by \cite{Mechtley2016} constitutes the best comparison to the study presented here, since they consider a sample of 19 high luminosity AGN at z$\sim$2, compared to the 32 high luminosity AGN at z$\sim$2 here. \cite{Mechtley2016} found a statistically insignificant enhancement in the probability of the fraction of AGN in mergers being greater than the fraction of controls of 0.78$\sigma$. Similar to figure 3 in \cite{Mechtley2016}, we show the probability density functions of the 32 z$\sim$2 AGN and their respective control galaxies in figure \ref{fig:PDFs}. Assuming the AGN in \cite{Mechtley2016} are intrinsically drawn from the same underlying population of AGN as those in this work, we can explore the likelihood of finding a statistically significant enhancement of the merger fraction in a smaller sample than presented here. Randomly drawing 19 AGN and 84 control galaxies from the sample \citep[matching to the number of AGN and controls in][]{Mechtley2016}, finding the statistical difference in merger fraction between the two populations, and repeating multiple times yields a wide (Gaussian) distribution of statistical significances. The probability that \cite{Mechtley2016} would find a significance of <1$\sigma$ is $\sim$24\%. Work to date has not found a statistically significant redshift evolution of AGN in mergers \citep{Gabor2009, Allevato2011, Cisternas2011, Schawinski2011, Treister2012, Bruce2015, Mechtley2016}; it will be interesting to see if this elevation persists in future large high redshift samples, since we are now possibly approaching large enough samples to detect such an enhancement.

\begin{figure}
	\includegraphics[width=\columnwidth]{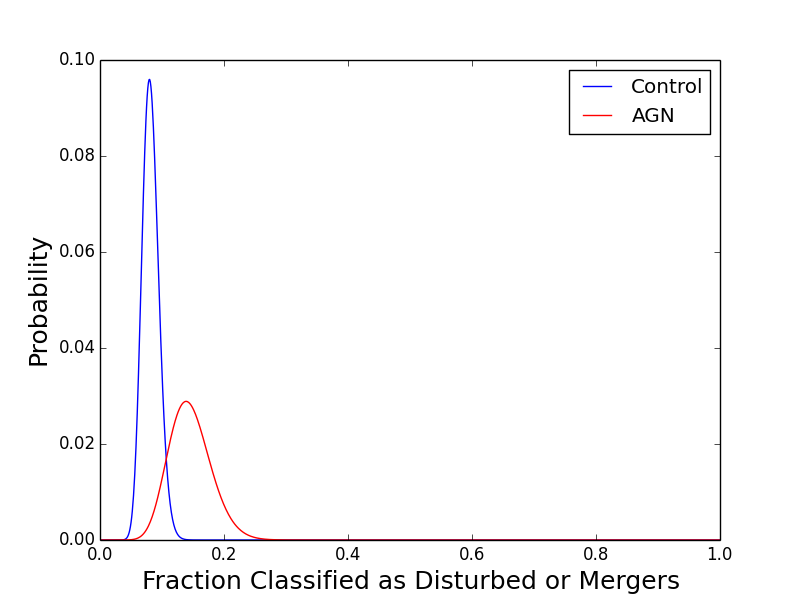}
    \caption{Probability density functions, calculated in accordance with \citet{Cameron2011}, for the fraction of AGN (wider red line) and controls (narrow blue line) classified as either mergers or disturbed.}
    \label{fig:PDFs}
\end{figure}

Overall AGN are not found more commonly in galaxies exhibiting signs of recent major mergers, but an excess of mergers is found in the highest redshift bin (z$\sim$2). No evidence is found in support of conjecture that the most luminous AGN must be fuelled predominantly by mergers, in contradiction to \citet{Treister2012} but in accordance with other studies \citep{Cisternas2011, Allevato2011, Karouzos2014, Mechtley2016, Villforth2017}. An enhancement of merger fraction in AGN hosts is found in the highest redshift bin, z$\sim$2, despite the results of other studies \citep{Gabor2009, Allevato2011, Cisternas2011, Schawinski2011, Bruce2015, Mechtley2016}. The majority of galaxies here are unlikely to have recently undergone a merger, given their disk-dominated light profiles \citep{Toomre1972, Barnes1988, DiMatteo2005, Cox2008, Gabor2009, Cisternas2011, Schawinski2011, Kocevski2012, Villforth2014, Villforth2017}.

\section{Conclusions}
\label{sec:Conclusions}

The aim of this study was to determine if a link exists between galaxy mergers and AGN activity, and whether this link evolves with redshift or AGN luminosity. High resolution HST images of 106 X-ray selected AGN and 486 host galaxy mass and redshift matched control galaxies from the COSMOS survey are analysed. Artificial AGN (stars), matched in apparent magnitude to the relevant AGN, are placed into control galaxies to make images equivalent. A new method for detecting disturbance in galaxies is developed: subtracting 2D photometric decomposition models from images (using \textsc{galfit}) and summing the remaining flux yields a numerical measure of merger features, since asymmetries and disturbances in the light profiles leave greater residual flux. The method correlates well with visual classifications. 

The key findings are summarised below, relating the residual flux and visual classification by three experts to AGN properties.

\hfill

i) Neither visual classification nor the residual flux method find evidence of greater fractions of AGN in merging galaxies, with similar fractions of merging control galaxies at all levels of disturbance. 

ii) Neither method finds evidence that mergers play a more dominant role at higher AGN luminosities, with similar fractions of merging control galaxies and AGN hosts at all luminosities. This suggests even the most luminous AGN are predominantly triggered secularly.

iii) Moderate evidence for redshift evolution of the role of mergers in AGN triggering is found for the first time, with a factor of $\sim$4 enhancement of the visually classified merger fraction in AGN hosts at z$\sim$2 relative to control galaxies, with 99\% confidence.  

iv) The fraction of galaxies exhibiting evidence of recent mergers in this sample is low ($\leq$20\%), suggesting that the majority of X-ray selected AGN at high luminosity are likely to be triggered alternatively, perhaps secularly or by minor mergers, at 0.5$\leq$z$\leq$2.2.
 
 \hfill

Although the single-filter approach to morphological analysis means that star forming regions will be over-sampled at high redshift as the rest frame shifts into the UV, by ensuring control galaxies are similar to AGN hosts like-for-like comparisons can still be made with validity. The residual flux technique is an effective way of picking out mergers in an automated way, and could be utilised on larger datasets, particularly with the aim of finding if only the most catastrophic mergers cause AGN of all luminosities. 

The detection of an enhancement of the incidence of mergers at z$\sim$2 is reported tentatively, given its absence in previous studies. Future work will have to use large samples (>30 luminous AGN) and find a similar enhancement of the merger fraction to lend credence to the results here.

\section*{Acknowledgements}

TH was supported by STFC grant ST/M503812/1 during the course of this work. V. W., J.M.-A., M.P.  and K.R. acknowledge support from the European Research Council Starting Grant SEDmorph (P.I. V. Wild). The table manipulation software, TOPCAT, was extremely useful at various points in this project \citep{Taylor2005}. This research made use of Astropy, a community-developed core Python package for Astronomy \citep{AstropyCollaboration}. The authors thank the anonymous referee for their enormously constructive comments. 




\bibliographystyle{mnras}

\bibliography{ResidualFluxPaperRef}




\newpage

\appendix
\section{}
\label{sec:Appendix}

\subsection{Residual Flux}
\label{subsec:ResidualFlux}

The primary question under investigation was whether the brightest AGN were in merging galaxies and how this changes with redshift. Merger signatures were quantified by subtracting \textsc{galfit} models from images and measuring the remaining flux as a proxy for morphological perturbation. Subtracting models from images leaves a profile consistent with noise if the fits are good; if the light profiles are discontinuous or asymmetric however, such as expected in merging systems, \textsc{galfit} will be unable to make good fits, leading to areas of over- or under-subtraction in the residuals. Summing the residual flux in the under-subtracted regions and the modulus of the residuals in over-subtracted regions, for values greater than the error on each pixel, gives a direct numerical measure of disturbance. This method was developed in part because at z=2 with L$_{\text{X,AGN}}$ $\sim$ 10\textsuperscript{45} erg s\textsuperscript{-1} visual classification becomes difficult in many cases, is highly subjective, and other automated methods were unlikely to be sensitive to the features searched for. In addition, this may be useful for future studies involving very large samples for quickly identifying those galaxies most likely to be mergers.

Methods in the literature for measuring the symmetry or internal disorder of a galaxy include the rotational asymmetry \citep{Shade1995, Rix1995, Abraham1996, Villforth2014}; the clumpiness parameter \citep{Isserstedt1986, Conselice2003}; the Gini index \citep{Lotz2004} or the shape asymmetry \citep{Pawlik2016}. The problem with applying these measures to the study of AGN lies in the low surface brightness of the features being searched for compared to the high intensity emission from the AGN. For example, the rotational asymmetry parameter is dominated by the bright central regions and not by extended tidal tails; the clumpiness parameter falls short if star forming regions are differently resolved at different redshifts; the shape asymmetry works exceptionally well at detecting low surface brightness features but is poorly suited to detecting multiple cores. The residual flux method shares in these challenges, but was born of an attempt to create a measure which simultaneously measures the internal disorder; the extended emission and the presence of multiple nuclei; and features which are otherwise outshone by the AGN, giving a continuous scale of disorder. Residual flux is not a measure that is directly comparable to other measures of disturbance such as those discussed above. In the future, the residual flux method may be utilised to pick out likely merger candidates for further inspection and analysis: since mergers are rare \citep{Bundy2009} gaining large samples of mergers can be challenging, particularly at high redshift where galaxies are most poorly resolved. 

Figure \ref{fig:Resids1D} shows the residual flux across the centre of a disturbed galaxy, exposing clearly a merger feature and demonstrating that merger features may very plausibly show up as residuals in fits: the more disturbed the galaxy the more residuals remain. With this motivation, the residual flux in annuli about the point source was calculated. The central circle with radius r\textsubscript{mask}=15 pixels was masked out because this region was PSF dominated, with much of the residual flux here being attributed to under-sampling of the PSF (see figure \ref{fig:Resids1D}). Through fitting to stars it was found that beyond 15 pixels this effect was largely mitigated, with the mean residual flux for stars with faint apparent magnitudes dropping close to zero. Annuli were at 30, 40 and 50 pixel radii in order to capture different regions of the galaxy ($\sim$ 7.5, 10 and 13 kpc radii at z=2, and is relatively constant over z=0.5-2.2). It was found that the inner 30 pixels (minus the central 15) were sufficient to qualitatively compare disturbance between galaxies, with broad results remaining the same for all annulus radii, independent of redshift.

High residual flux remaining after the model subtraction is highly suggestive of some non-equilibrium state \citep[a galaxy in equilibrium will relax into a symmetric state;][]{Robertson2006}: e.g. when galaxy morphologies are perturbed by major mergers. Residual flux is not just an on-off switch signifying mergers however, but a sliding scale of disturbance. This may help to distinguish between major and minor mergers, or recent and old mergers. The more morphological disturbance the less smooth the light profile and the more residual flux remaining post model-subtraction. Merger features are expected to fade away on relatively short timescales \citep[totally faded within $\sim$1Gyr, typically less][]{Larson1978, Kennicutt1987, Lotz2010b, Lotz2010a, Ellison2013} and so moderate values of residual flux may indicate some relaxing state after a merger, or some other form of disturbance such as minor mergers, while the highest values of residual flux should belong to major mergers. Thus, this has the power to differentiate somewhat between triggering mechanisms, or possibly to constrain any possible time delay between mergers and AGN activity. Occasionally host galaxies are only marginally resolved. This can result in an incomplete \textsc{galfit} model and an overestimation of the residual flux as due to disturbance. In these cases the residual flux is generally low, as a bright underlying galaxy will be fit better by \textsc{galfit}, but visual classification by 3 expert classifiers was used to identify marginally detected hosts, so that they could be excluded from analysis to determine the significance of this factor (see section \ref{subsec:VisualClassification}).

 Figure \ref{fig:ResidsClassCorrelation} shows the residual flux for the different morphological categories. There is no particular difference between classifications of undisturbed galaxies with or without residuals, suggesting the residual flux method is similarly good at picking out true disturbance as humans. The residual flux is not heavily influenced by non-axisymmetric features such as spiral arms: omitting objects flagged by visual classifiers as having such features did not influence results. The mean residual flux rises for disturbed and again for merger classes: on average this method is picking out similar features to humans. Although there is considerable scatter within any classification, a galaxy with residual flux > 2 has a >50\% chance of being classified by humans as a merger or disturbed, despite these constituting <20\% of the sample. This highlights the power of this method for identifying large samples of mergers in future studies.

\begin{figure}
	\centering
	\includegraphics[width=\columnwidth]{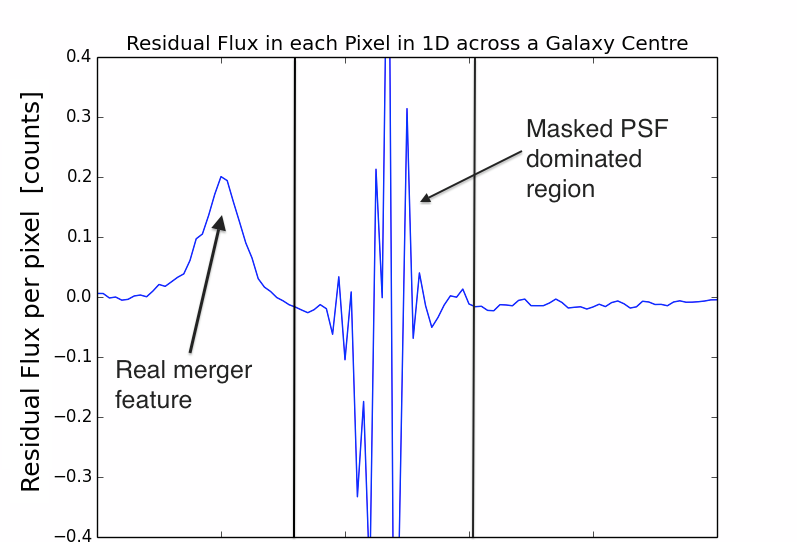}
	\includegraphics[width=0.88\columnwidth]{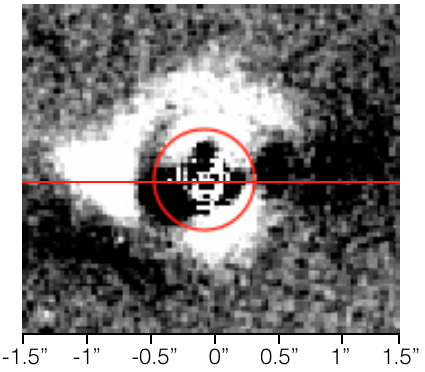}
    \caption{Upper panel: residual flux in a one dimensional horizontal line across the centre of the disturbed AGN host in the lower panel. Vertical lines denote the region masked out due to diffraction spikes in PSF residuals, seen as a mosaic-like region in the centre below, while a merger feature is highlighted at around -1". The image below is aligned with the upper panel and the angular distribution on the sky is displayed in arcseconds, with the horizontal line and masked out region, with r\textsubscript{mask} = 15 pixels, denoted by the red line and circle. X-ray ID=5133, z=0.66}
    \label{fig:Resids1D}
\end{figure}

\begin{figure}
	\includegraphics[width=\columnwidth]{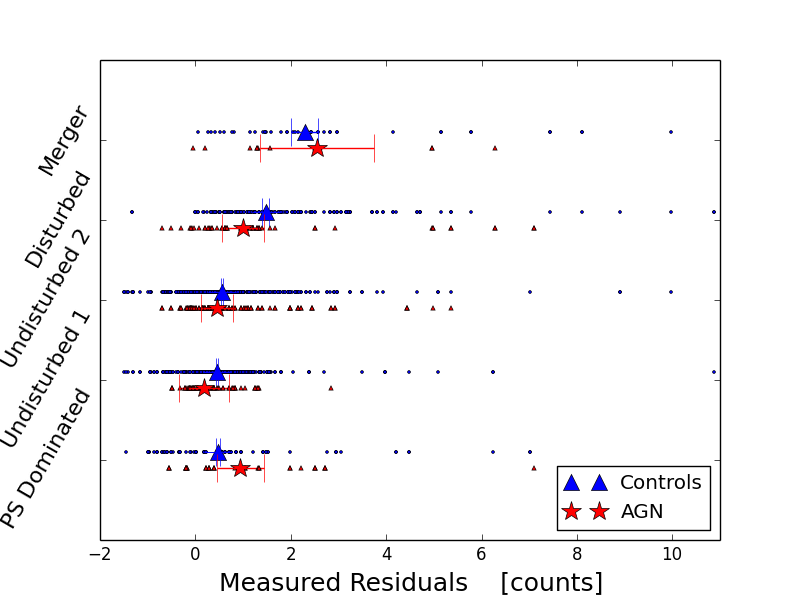}
    \caption{Human visual classification of morphology vs residual flux. Classifications are point source dominated galaxies (PS dominated); undisturbed without residuals (undisturbed 1); undisturbed with residuals (undisturbed 2); disturbed galaxies and clear mergers. Small red and blue points are the individual galaxy classifications while the big red stars and blue triangles are median values of residual flux for each classification (AGN and Controls respectively). Errors are found by bootstrap resampling. 1/3 of the data is randomly sampled (and replaced) and the mean found; this is repeated 1000 times and the standard deviation on those mean values provides the final error.}
    \label{fig:ResidsClassCorrelation}
\end{figure}

\subsection{Accounting for Potential Bias}
\label{subsec:Bias}

To ensure that results were not biased significant effort was made to control for factors that may influence the measurement of the residual flux: poor point source subtraction; brighter galaxies having higher residual flux just because they are brighter to start with; and relations between residual flux and the signal to noise ratio (SNR). The most important consideration is for any biases to be equal between AGN hosts and controls, since enhancement of one over the other would be the significant discovery.

First, to control for poor point source subtraction, point sources were fit in \textsc{galfit} to the 26 stars used as artificial point sources in the control galaxy group (section \ref{subsec:Controls}) and subtracted from the images. Figure \ref{fig:StarResids} shows their residual flux as a function of apparent magnitude. Brighter point sources have greater residual fluxes, unrelated to the galaxy morphology. The tight correlation between residual flux and magnitude can, however, be used to correct for point source light contamination. AGN and artificial point sources were matched to the closest magnitude star and the residual flux of that star was subtracted from the residual flux value obtained for the AGN+host or artificial point source+galaxy system, meaning any remaining residual flux should be due to the galaxy itself and not poor point source subtraction. Since figure \ref{fig:StarResids} shows how the trend of residual flux vs magnitude flattens at magnitudes fainter than $\sim$21, and there is no reason to expect this trend to change at even fainter magnitudes, using the faintest star as a surrogate for even fainter AGN (section \ref{subsec:Controls}) should not bias the results. We chose to correct for point source light contamination by subtracting residual flux, since there was such a tight relationship between magnitude and residual flux, instead of scaling down point source flux to exactly match the AGN, since this could have introduced new and unknown sources of bias. The brightest stars were comparable to the brightest AGN. Some variation in the PSF over the field may be expected, however no trend between residual flux and distance from the centre of the chip was found, and no obvious asymmetries were detected in the majority of the stars from visual inspection. This suggests the overwhelmingly dominant factor influencing the residual flux from the point source is its magnitude.

\begin{figure}
	\includegraphics[width=\columnwidth]{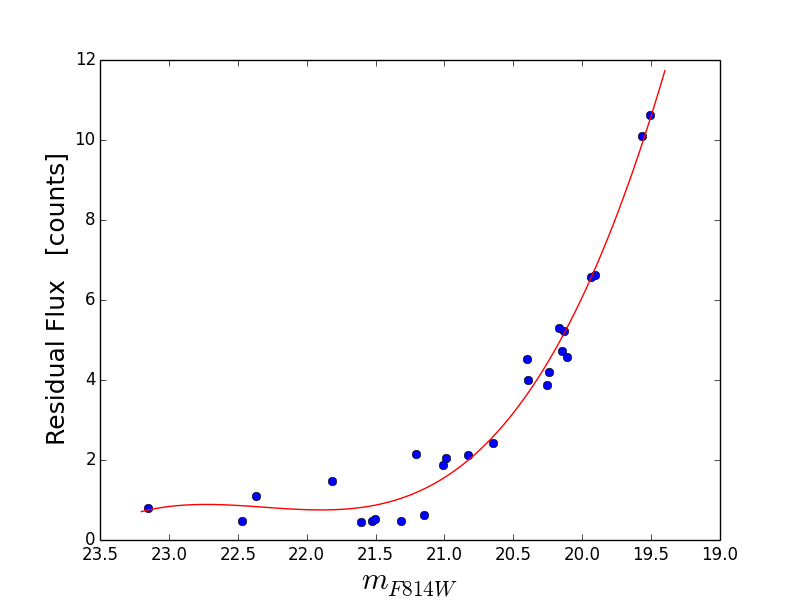}
    \caption{Residual flux of stars in annulus from 15-30 pixels as a function of magnitude. The tight correlation can be used to control for poor point source subtraction.}
    \label{fig:StarResids}
\end{figure}

Second, no correlation between control galaxy stellar masses and residual flux was found. Mass was used as a proxy for magnitude as a check that is independent of the \textsc{galfit} models, since model magnitudes are clearly degenerate with the residual flux, and F814W magnitudes are contaminated with AGN light for the AGN hosts. Since we found no correlation between stellar mass and residual flux this strongly suggests the brightness of the host is not a significant factor in producing high residual fluxes. 

Finally, since residual flux of a pixel is counted only if its value is greater than the error on that pixel, some correlation between residual flux and the SNR may be expected. Indeed, a correlation between SNR and residual flux was found. AGN and control galaxies are found to be equivalent however, so this will not cause any kind of enhancement in one group over the other. Because in the control group two images have been added together (galaxy and star) the noise is greater and the sky flux is correspondingly higher. For this reason 1$\sigma$ is the correct choice above which to sum flux, since errors take account of the addition of images by summing the image errors in quadrature, before subtracting the point source contribution defined by figure \ref{fig:StarResids}.

Figure \ref{fig:Resids5} shows all galaxies with residual flux > 5 counts, most demonstrating high degrees of disturbance; five galaxies were removed due to non-physical features in the images such as the light from overexposed nearby stars bleeding over the galaxy image. The highest residual flux galaxies appear to have the highest degrees of disturbance: this technique picks these features out and identifies strong merger candidates, in addition to the increased statistical likelihood of a galaxy being disturbed if residual flux is high.

\begin{figure*}
	\includegraphics[width=0.24\textwidth]{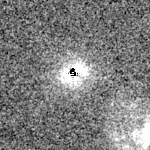}
	\includegraphics[width=0.24\textwidth]{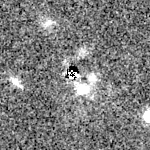}
	\includegraphics[width=0.24\textwidth]{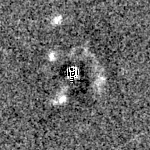}
	\includegraphics[width=0.24\textwidth]{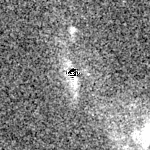}
	\includegraphics[width=0.24\textwidth]{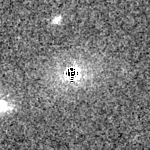}
	\includegraphics[width=0.24\textwidth]{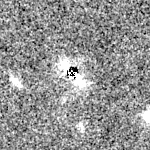}
	\includegraphics[width=0.24\textwidth]{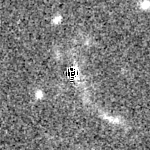}
	\includegraphics[width=0.24\textwidth]{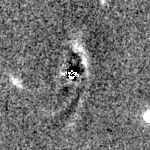}
	\includegraphics[width=0.24\textwidth]{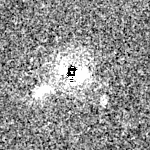}
	\includegraphics[width=0.24\textwidth]{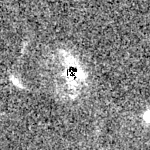}
	\includegraphics[width=0.24\textwidth]{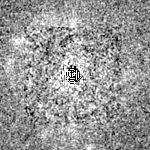}
	\includegraphics[width=0.24\textwidth]{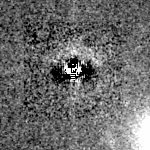}
	\includegraphics[width=0.24\textwidth]{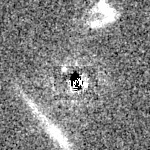}
	\includegraphics[width=0.24\textwidth]{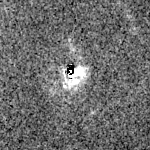}
	\includegraphics[width=0.24\textwidth]{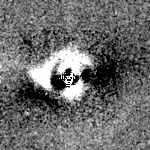}
	\includegraphics[width=0.24\textwidth]{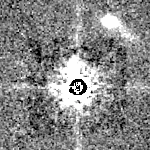}
    \caption{All galaxies with residual flux > 5 counts; high degrees of disturbance are seen in almost all cases. The first 11 galaxies (counting left to right, top to bottom and starting in the top left) are controls and the last 5 are AGN hosts. 5 Galaxies (1 AGN and 4 control) were removed due to image defects; see text for details.}
    \label{fig:Resids5}
\end{figure*}


\bsp	
\label{lastpage}
\end{document}